\documentstyle[12pt,agums]{article}
\lefthead{FRISCH}
\righthead{GALACTIC ENVIRONMENT OF THE SUN}
\received{April~9,~1999}
\revised{May~26,~1999}
\accepted{May ~26,~1999}
\paperid{}
\cpright{AGU}{1999}
\ccc{0148-0227/99/1999JA900238\$09.00}
\authoraddr{P.~C. Frisch, 
Department of Astronomy and Astrophysics,
5640 South Ellis Avenue,
University of Chicago, Chicago, Illinois  60637
(e-mail: frisch@oddjob.uchicago.edu)}

\begin{document}

%% ------------------------------------------------------ %%
%
%%  TITLE
%
%% ------------------------------------------------------ %%

\title{The galactic environment of the Sun}

%% ------------------------------------------------------ %%
%
%%  AUTHOR NAMES, AFFILIATIONS, and ALTERNATE AFFILIATIONS
%
%% ------------------------------------------------------ %%

\author{Priscilla C. Frisch \altaffilmark{1} }

\affil{Department of Astronomy and Astrophysics,
University of Chicago, Chicago, Illinois}

\begin{abstract}
The interstellar cloud surrounding the solar system regulates the
galactic environment of the Sun and constrains the
physical characteristics of the interplanetary medium.
This paper compares interstellar dust grain properties observed within the solar system 
with dust properties inferred from observations of the cloud surrounding the solar
system.  Properties of diffuse clouds in the solar vicinity are
discussed to gain insight into the properties of the diffuse cloud complex
flowing past the Sun.
Evidence is presented for changes in the galactic environment of the Sun within
the next 10$^4$--10$^6$ years.
The combined history of changes in the interstellar environment of the Sun,
and solar activity cycles, will be recorded in the variability of the ratio of
large- to medium-sized interstellar
dust grains deposited onto geologically inert surfaces.  Combining
data from lunar core samples in the inner and outer solar system will
assist in disentangling these two effects.
\end{abstract}

\begin{article}
\section{Introduction}\label{intro}

The interstellar cloud surrounding the solar system regulates the
galactic environment of the Sun and constrains the
physical characteristics of the interplanetary medium enveloping
the planets.  In addition, the daughter products of the
interaction between the solar wind and the interstellar cloud surrounding 
the solar system, when compared with astronomical data,
provide a unique window on the chemical evolution of our galactic neighborhood,
reveal the fundamental processes that link the interplanetary environment to
the galactic environment of the Sun,
and constrain the history and physical properties of one sample of a diffuse interstellar cloud.  
The combined history of variability in the solar galactic environment 
and activity cycles will be recorded in the variability of the ratio of 
large- to medium-sized interstellar
dust grains deposited onto geologically inert surfaces within the solar system
and, if this record could be accessed (and has not been destroyed), will witness the journey of the
Sun through space.  Oort cloud comets will sweep up pristine interstellar dust
outside of the heliosphere.

The first effort to identify the interstellar cloud feeding gas and dust 
into the solar system found a difference of several kilometers per second 
between the velocity of the nearest interstellar gas in the upwind direction 
and interstellar gas inside the solar system identified spectrally
by H$^{\rm \circ}$ Lyman -$\alpha$ backscatter radiation [{\it Adams and 
Frisch}, 1977].
We now know that this discrepancy is due to a velocity gradient across the 
complex of interstellar clouds surrounding the solar system (section \ref{velocity}),
possibly reflecting its fractal nature (section \ref{variable}).
On the basis of evidence of shock-front destruction of dust grains in nearby interstellar
material, and the fact that the bulk velocity of the local interstellar
matter (LISM) is consistent with 
the outflow of interstellar matter (ISM) from the 
Loop I superbubble (a curved radio-emission feature seen in the galactic center hemisphere of the
sky; section \ref{variable}), 
I concluded that interstellar matter inside and immediately outside
of the solar system
was part of an expanding superbubble shell around Loop I [{\it Frisch}, 1981, 1996].
On the basis of detailed models of star formation in the Scorpius-Centaurus
Association, the superbubble shell formed with the creation of the
Scorpius subgroup appears to have expanded to the location of the Sun
[{\it De Geus}, 1992; {\it Frisch}, 1995].  Alternative explanations place the
Sun at the boundary of two merged superbubbles [e.g., {\it Egger}, 1998], but this
alternative fails to explain the flow of nearby ISM from the Scorpius-Centaurus
region.  The conclusion that enhanced gas-phase abundances of
refractory elements (Ca, Fe, Mg) in nearby interstellar clouds is caused by
grain destruction in shock fronts is now supported by detailed calculations
(see section \ref{licdust}).
The recent successes of the
Ulysses and Galileo dust detectors 
(see following section)
in observing the mass distribution
of ``typical'' interstellar dust grains have now opened new possibilities
for understanding interstellar dust.

The interstellar cloud in which the solar system is currently
embedded, also known as the local interstellar cloud
(LIC, see section \ref{velocity}), is low density and warm 
($n$(H$^\circ$+H$^+$)$\sim$0.3 cm$^{-3}$, {\it T}$\sim$6900 K).
The physical properties of the cloud around the solar system, including
temperature, electron density and magnetic field strength, have been summarized
elsewhere [e.g., {\it Frisch}, 1990; {\it Lallement}, 1998] and are listed in 
\callout{Table~1}.
The interplanetary environment is subject to the physical conditions of
the cloud surrounding the solar system, and currently $\sim$98\% of
the diffuse gas within the heliosphere (by number) is interstellar, with the
densities of the solar wind and interstellar matter equal near the orbit of Jupiter.

Today, the solar wind effectively prevents most interstellar atoms,
low-energy cosmic rays (energy/nucleon $\leq$300 MeV), and small dust
grains from reaching the inner solar system.  Over the history
of the solar system the outer
planets, comets, and Kuiper Belt objects are more likely to be immersed in pristine interstellar
material than are inner objects.  
It has been shown that if the density of the surrounding cloud increases moderately,
both the properties of the heliosphere and 
the density of H$^{\rm \circ}$ and He$^{\rm \circ}$ in the inner solar system
change dramatically.  For instance, if the density 
of the surrounding cloud is increased to $n_{\rm H}$=10 atoms cm$^{-3}$
(above the current value of $n_{\rm H}$$\sim$0.3 atoms cm$^{-3}$; Table 1)
the termination shock contracts to 10--14 AU upstream and is 
highly dynamical, while the density of H$^{\rm \circ}$ increases
to $\sim$2 cm$^{-3}$ at 1 AU [{\it Zank and Frisch}, 1999].  
Arguments about the fractal nature of the interstellar medium, within the
boundaries set by the global structure imposed by forces shaping 
cloud morphology, predict maximum density contrasts of $\sim$10$^4$ 
within a cloud complex [{\it Elmegreen}, 1998].  
If similar density contrasts are present in the cloud complex 
sweeping past the Sun, the interplanetary medium at Earth orbit
potentially could change dramatically on short timescales.

\section{Comparison of Gas and Dust in the 
Local Interstellar Cloud\label{licdust}}

Comparisons of interstellar dust grains
observed within the heliosphere by the Ulysses and Galileo spacecraft 
[e.g., {\it Baguhl et al.}, 1996; {\it Landgraf}, this issue; {\it Landgraf et al.}, this issue]
with interstellar dust properties inferred from observations of gas in the LISM
[{\it Frisch et al}, 1999] 
provide a unique tool that can be used to probe 
the origin and evolution of interstellar dust grains,
to constrain galactic chemical evolution, and
to evaluate the behavior of interstellar dust in the inner heliosphere
were the Sun to encounter a denser interstellar cloud.
Interstellar dust grain models have been formulated to predict the extinction 
and polarization of optical starlight by interstellar dust grains
and do not tightly constrain the numbers and properties of large 
dust grains (radius $a_{\rm gr}\geq0.3$ $\mu$m) [e.g., {\it Mathis}, this issue;
{\it Witt}, this issue;
{\it Kim et al.}, 1994; {\it Li and Greenberg}, 1997), so that spacecraft observations of interstellar dust 
with $a_{\rm gr}\geq0.3$ $\mu$m add new information to our understanding of 
interstellar dust.  
The bulk of the discussion in this section is taken from 
{\it Frisch et al.}, [1999] (hereinafter referred to as F99).

The population of dust grains found in interstellar space
includes grains with a range of mass, charge, size and composition
properties [e.g., see {\it Witt}, this issue; {\it Mathis}, this issue ]
and each population interacts differently with the
heliosphere.  
The derivation of the mass distribution of inflowing interstellar dust grains
from in situ spacecraft observations relies on modeling the
interaction of these grains with the heliosphere and heliosheath
regions.
Very small ($a_{\rm gr}<$0.01 $\mu$m) charged interstellar dust grains
are excluded by Lorentz force interactions
with the compressed interstellar magnetic field (when present) in the heliosheath
region between the heliopause (HP) and bow shock (F99).
(In this paper, all conversions between grain mass and radii are made
assuming spherical grains with density 2.5 g cm$^{-3}$, 
giving $m_{\rm gr}\sim10.5 ~ a_{\rm gr}^3$, where $a_{\rm gr}$ is measured in centimeters.)
Grains with mass $m_{\rm gr}<10^{-14}$ g ($a_{\rm gr}\leq$0.1 $\mu$m)
are excluded by Lorentz force interactions with the solar wind,
while the spatial distribution within the heliosphere of somewhat larger grains is modulated by the
solar wind cycle [see {\it Landgraf}, this issue; {\it Mann and Kimura}, this issue].
The strength of the
interstellar magnetic field in the interstellar cloud surrounding the
solar system is not known, and if the field strength is zero, the
Lorentz force deflection of very small grains in the ``hydrogen wall'' 
pileup will not be effective.  In this case, alternative filtering
mechanisms within the heliosheath and outer heliosphere regions 
occur, but the distribution of small dust grains in the outer heliosphere
and heliosheath regions will differ distinctively from the case where the 
interstellar magnetic field strength is nonzero.

Comparisons between the gas-to-dust mass ratio value
\mbox{($R_{\rm g/d}$)} derived from the in situ
versus astronomical data provide insight into
the reference abundance for the interstellar medium and grain history.
The in situ value 
yields $R_{\rm g/d}$=94$^{+46}_{-38}$, from comparisons of 
the mass of interstellar dust grains observed by
the Ulysses and Galileo satellites with the gas mass densities
for the surrounding interstellar cloud.  
%When the total mass of the grains detected in situ
%is compared to the LIC density, a gas-to-dust mass ratio of
%$R_{\rm g/d}$=94$^{+46}_{-38}$ is found.  
However, since small grains
($a_{\rm gr}\leq$0.1 $\mu$m) are excluded from the inner heliosphere where
the data were acquired, the true value of $R_{\rm g/d}$ for the
cloud material flowing into the heliosphere should be smaller
than this value.  The exception to this conclusion would be if small and
large dust grains are not mixed homogeneously in space, 
because of grain processing within molecular clouds,
in the cloud disruption process,  and in diffuse gas
[{\it Witt}, this issue].

%The life cycle of interstellar dust grains includes injection of seed-grains
%by supernova and the winds of active giant branch stars (AGB).  
%(??????)  Most of the dust mass, however, is believed to form by the 
%depletion of gas-phase atoms to form mantles on small seed grains in 
%dense molecular clouds [e.g., {\it Tielens}, 1998, {\it Dwek}, 1998).

The gas-to-dust mass ratio can also be found from observations of
absorption lines in nearby stars.  Grain mass builds up in molecular
clouds by the accretion of gas atoms onto the surfaces of small grains
(which have most of the surface area of interstellar grains and are
negatively charged so that ions will stick) 
[e.g., {\it Weingartner and Draine}, 1999].
The dust mass of a cloud can then be determined by
applying a ``missing mass'' argument, which is based on the assumption that 
the sum of the
atoms of a given element in the combined gas and dust phase 
is equal to the ``reference abundance'' of that atom.   This
logic is developed and applied to observations of the cloud
surrounding the solar system, in order to determine the 
gas-to-dust ratio from astronomical data (F99).  In principle, comparisons
between the spacecraft and astronomical results should allow
the reference abundance of the gas to be removed as a variable
if this logic is correct.  In reality, the disruption of the
molecular cloud that is parent to the cloud surrounding the
solar system occurred millions of years ago, and the application
of this argument requires the assumption that gas and dust in the
cloud have remained closely coupled over this time (by collisions
and coupling to an embedded magnetic field)
with no new grains captured in the expanding material.

The gas-to-dust mass ratio in the LIC cloud, based on observations of interstellar
absorption lines in the ultraviolet region of the spectrum
and observed in the spectrum of the star $\epsilon$ CMa, is
larger than the ratio based on the spacecraft data, with the exact value depending on
the assumed reference abundance, which is unknown.
If solar reference abundances are 
assumed, $R_{\rm g/d}$=427$^{+72}_{-207}$ for the LIC, while
$R_{\rm g/d}$=551$^{+61}_{-251}$ if B-star reference abundances are
assumed.  In either case, the gas-to-dust mass ratios derived from
in situ versus astronomical observations of the LIC differ by
large amounts.  (The quoted uncertainties on the astronomical
determinations represent 2$\sigma$ or better uncertainties (see F99).)
These differences suggest that the interstellar dust grains
observed in situ include a population which has not
exchanged dust with the LIC -gas, thereby invalidating the 
missing-mass argument.
This extra population may be dust particles from any source,
including circumstellar material, which has a separate history
from the LIC.  For example, any population of dust grains
captured by a magnetic field embedded in the moving LIC cloud 
(which is moving through space at $\sim$19 km s$^{-1}$)
would not have exchanged atoms with the LIC and therefore
would enrich the dust population and invalidate the missing-mass
argument.  Alternatively, the metallicity of the LISM may be inhomogeneous.

The classic view for ISM reference abundances in the solar 
neighborhood is that solar abundances apply
(even though the Sun has long since decoupled from its parent protosolar
nebula).  More recently, it has been shown that a wide range of abundances 
are found in solar type stars [{\it Edvardsson et al.}, 1993], and that
reference abundances for both ISM and 
young B-stars appear to be metal-poor in comparison with solar values
[{\it Snow}, this issue; {\it Mathis}, this issue).
If the reference abundances for metals are reduced, that leaves
intrinsically less mass for the formation of interstellar dust
grains, in comparison with assumed solar abundances, when missing-mass 
arguments are invoked [e.g., {\it Savage and Sembach}, 1996] (hereinafter
referred to as SS).
The dust grain core-mantle composition can be
inferred by assuming that the grain core composition is the same as the grain composition found 
for the most weakly depleted clouds observed,
which are typically in the direction of halo stars. 
After the core composition is subtracted from the grain mass,
the remaining material in the grains 
is then attributed to the grain mantles (SS).
This logic depends on the missing-mass argument, which effectively 
argues that the total mass of any seed grains injected from either 
asymptotic giant branch stars (AGBs)
or supernova is negligible over the lifetime of the grain population.  

Over 80\% of refractory elements Mg and Fe
in the interstellar medium are 
condensed onto dust grains, and the refractories found in the gas
phase result from the destruction of dust grains by sputtering in 
shock fronts [{\it Jones et al.}, 1996].   
The high abundances of refractory elements in shocked clouds indicates that 
sputtering of refractory-rich small grains occurs in shocks
(although the small grain population also reforms from the
shattering of large grains) [{\it Jones et al.}, 1996].
The gas-phase abundances of Fe, Mg, and Si in the velocity component
corresponding to the LIC velocity projected
toward $\epsilon$ CMa, which is believed to be the
interstellar cloud surrounding
the solar system, indicate grain destruction of
the parent dust grains by a shock of
velocity of 100--200 km s$^{-1}$.  The relatively
good correlation between the Mg and Fe column densities in the LISM gas
indicates Fe and Mg originate from destruction of the same dust component.
(The ``column density'' of an element is the number of atoms seen
in a column that has a cross -section of 1 cm$^{-2}$ and a length equal to the
distance of the background star.  The units of column density are per
square centimeter.)
In the LIC, comparison between gas-phase abundances (using
$\epsilon$ CMa data) and grain core (SS) models indicates that the destroyed mantle
material was silicon rich in comparison with the cores.
The lack of Si in LIC dust indicates grain mantles have
been mostly eroded.  
The enhanced abundance of refractories in the LIC are also consistent with
cosmic -ray acceleration models which require small refractory-rich dust grains ($a_{\rm gr}\sim0.1$ $\mu$m) to be accelerated and
destroyed by shocks in order to provide the initial injection of refractory 
atoms into shocks for acceleration up to cosmic -ray energies 
[e.g., {\it Ellison et al.}, 1997].
(Volatiles are accelerated to cosmic -ray energies directly from the gas phase.)

If a significant fraction of the LIC dust mass is contained in small 
and very small dust grains, such as found elsewhere in the ISM, 
then the discrepancy between the $R_{\rm g/d}$ value determined from in situ
and astronomical data is exacerbated.  Alternatively, it is possible 
that the shock front which destroyed the small grains, and returned the refractory 
elements to the gas -phase, entirely removed these small grains.  
Grains with radii $a_{\rm gr}$$>$1 $\mu$m survive in interstellar
shocks because of the underabundance of particles capable of fragmenting 
them [e.g., {\it Jones et al.}, 1996].  This scenario would explain the
relatively large mass of interstellar grains observed by the spacecraft.

\section{The LIC Among Diffuse Clouds\label{diffuse}}

% The distribution of nearby interstellar gas around the location of the
% Sun is asymmetric, with an order of magnitude more ISM found in the
% galactic center hemisphere than in the anti-center hemisphere.  
% Fig. \ref{LISM} shows the sky divided into two regions:  regions
% where the total column density of material within
% 30 pc is greater than 10$^{18}$ cm$^{-2}$, and regions where it
% is not.  The top figure gives the distribution in galactic coordinates, while the
% bottom shows the distribution in ecliptic coordinates.  

The interstellar cloud surrounding the solar system is
part of a complex of diffuse interstellar clouds flowing past the
Sun, and an analogous system should be present elsewhere in the galaxy.
The identification of a LIC-type
diffuse cloud toward an external star is more likely in regions which either have been disturbed by star formation and supernova activity
or which show high-velocity clouds from infalling halo gas, causing
low column density components to be separated
in velocity from other clouds in the sight line.  (High spectral resolution observations are required
to resolve individual velocity components in complex sight lines,
which in turn mandates a relatively bright background star with
minimal reddening by foreground interstellar dust.)  
Such rapidly moving clouds have long been known to show enhanced
abundances of refractory elements [{\it Routly and Spitzer}, 1952], a property
shared by the LIC [e.g., {\it Frisch} 1981; {\it Bertin et al.}, 1993; 
{\it Gry and Dupin}, 1996; {\it Gry et al.}, 1995].

The identification of a cloud is
based on fitting an absorption line formed by an ensemble of interstellar 
atoms (or ions) in the cloud 
with a Maxwellian velocity distribution, characterized by
a kinetic temperature and turbulent velocity.
%(typically about 1.00--1.5 km s$^{-1}$ in the LISM, e.g., Linsky et al., 1995).
Improvements in instrumental spectral resolution when observing
cold clouds always reveal
additional clouds unresolved at lower spectral resolutions.
(For instance, at a resolution of 0.5 km s$^{-1}$, $\sim$40\% 
of the clouds contributing to Na$^\circ$ absorption
lines are missed because of blending with adjacent
velocity components [{\it Welty et al.}, 1993].)  In warm diffuse gas this
may not be the case since the
velocity dispersion increases and
theoretical line widths are about 8 times as wide as in the
cold gas.  Low column density velocity components are always difficult to
identify, unless they are separated in velocity from adjacent components
or are seen in a nearby star.

LIC-type clouds were originally identified as high-velocity clouds
in halo stars.
For example, the halo star HD 215733 [{\it Fitzpatrick and Spitzer}, 1997] shows
warm and diffuse clouds.  Component 5, at --54 km s$^{-1}$, has $T$=7000$\pm4100$ K 
and log $N$(H$^\circ$)=19.63 cm$^{-2}$.
The electron density is $n$({\rm e$^-$})=0.05 cm$^{-3}$, 
based on the ionization equilibrium of Ca$^+$.
If the electron density is instead found from the collisional excitation of C$^+$, {\it n}(e$^-$)=0.0014--0.095
cm$^{-3}$ (depending on the detailed assumptions).
In this component, the depletion of Mg is comparable to depletions in 
the LIC toward the stars $\alpha$ Aur, $\alpha$CMi, and
GD191-B2B (e.g., F99), however the total column density is larger than
the LIC value.
Component 2, at --83 km s$^{-1}$, has a kinetic temperature 
comparable to the LIC, $T$=6000$^{+5000}_{-3100}$ K 
and log $N$(H$^\circ$))=18.40 cm$^{-2}$.
Many absorption lines are too weak to be visible in this weaker component.  
The LIC has less Fe depletion than either components 2 or 5,
indicating the LIC is one of the least depleted
sight lines [e.g., {\it Gry and Dupin}, 1996; {\it Dupin}, 1998].

The group of diffuse cloud components found in front of 23 Orionis
samples a region disturbed by an expanding
superbubble shell (although it is younger than the $\sim$4 million
year old shell near the Sun).  
A recent study of this sight line by {\it Welty et al.} [1999]
presents puzzling contradictions in our understanding of diffuse
clouds.
Two cloud complexes dominate the sight line:  the ``weak low velocity'' (WLV)
and the ``strong low velocity'' (SLV) groups.  
The column densities of the WLV and SLV groups are 
log $N$(H$^{\rm o}$)=19.61 cm$^{-2}$
and log $N$(H$^{\rm o}$)=20.71 cm$^{-2}$, respectively.
The WLV cloud complex yields insight into LIC-type clouds,
although it is cooler (possible temperature $\sim$3000 K),
higher column density, and more depleted than the LIC.

{\it Welty et al.} [1999] have shown that no unique electron density can be
derived for these clouds, with different methods yielding different values.
For the WLV complex, the ionization equilibrium for C, Na, and Mg
gives values ranging from {\it n}(e$^-$)=0.11 to 0.22 cm$^{-3}$ (for {\it T}=3000 K).
The properties of each individual component within the
WLV complex vary, preventing the determination of {\it n}(H$^\circ$),
and intrinsically weak spectral features cannot be observed in this 
component because of low column densities.
The higher column density SLV cloud complex is not 
a good model for the LIC but illustrates the limitations of our understanding.
The SLV is denser ($n_{\rm H}$$\sim$10--15 cm$^{-3}$) than the LIC,
and $\sim$1\% ionized, with less than 1\% of the gas in the 
form of molecular hydrogen.   
For the SLV, evaluating the ionization equilibrium
of 12 separate elements yields electron density values ranging
from {\it n}(e$^-$)=0.04 to 0.95 cm$^{-3}$.
The puzzling result that a consistent electron density is not found from 
different methods
indicates either that we do not yet understand the
physics of diffuse clouds or that atomic or rate constants are wrong.
On the basis of observations of the 23 Orionis sight line,
LIC-type clouds are not expected to have molecular material.
%Molecular hydrogen is not observed in the WLV component.  
%($n_{\rm H}$$\sim$10--15 cm$^{-3}$)
%In the cold and denser SLV component, less than 1\% of the gas is in the form of molecular hydrogen.
%The other observed molecules in the SLV component (CH, CH$^+$, CN)
%are at least 5 orders of magnitude less abundant than H$_2$.

%no, as the H_2 suggests that T is of order 100 K for the SLV gas (any reason
%to suspect that?)  For the WLV gas, the C I fine-structure ratios did not
%permit an estimate for n_H, as the individual comps seem to have different 
%properties.

\section{Time-Variable Galactic Environment of the Sun\label{variable}}

The galactic environment of the Sun changes with time.  In the 
rest velocity frame defined by the average motion of 
cool stars in our neighborhood of the galaxy 
(the ``local standard of rest,'' or LSR), nearby interstellar material is sweeping past the
Sun from the galactic center hemisphere.
%The space motions of the Sun and surrounding cloud complex are nearly perpendicular
Hence the solar environment during the next few years will be
dominated by the physical properties of nearby gas in
this hemisphere of the sky.

The LSR space motion of the LIC cloud can be found by removing the 
vector motion of the Sun through the LSR.  
The solar motion has been rederived recently using
precise astrometric data from the Hipparcos satellite, giving a new 
solar velocity of 13.4 km s$^{-1}$ toward galactic longitude and
latitude 28$^\circ$, +32$^\circ$ [{\it Dehnen and Binney}, 1998].  This new value represents a significant
change from the previous value (see footnotes to Table 1) and is not skewed 
by velocities of very young stars (which may retain the motion of
the parent molecular cloud) or very old stars (which lag galactic
rotation).

The morphology, flow direction, and dust grain destruction evident in the LISM
suggest that the cloud complex is part of a superbubble shell
associated with star -formation in the Scorpius-Ophiuchus Association
[{\it Frisch}, 1995] (hereinafter referred to as F95).
Within the global morphological structure set by the boundary 
conditions on a given volume of interstellar material,
the ISM is fractal with a fractal dimension
characteristic of turbulence [{\it Elmegreen}, 1998].   The fractal
structure allows density contrasts over short spatial scales.

The galactic environment of the Sun over the past several $\sim$10$^6$
years has been dominated by two types of material, both of which
should have yielded a heliosphere with about the same dimensions.
Within the past $\sim$10$^5$
years the Sun emerged from the hot low-density interior of 
the Local Bubble and entered the cloud complex now sweeping past the Sun
from a direction near the central region of the expanding
Loop I superbubble.  The absence of neutrals within the Local Bubble, combined with
high temperatures, indicates the heliosphere was large in that region,
providing solar wind characteristics were similar to today.
%(heliosphere radius $\sim$ 130 AU, Frisch 1998).
More recently, the Sun was probably in  the
warm ``blue-shifted'' cloud seen toward Sirius
and $\epsilon$ CMa, although the three-dimensional space trajectory of
this cloud is unknown.  These two regions have pressure
characteristics yielding
heliosphere radii of $\sim$130 and $\sim$120 AU, providing
the solar wind was unchanged from today [{\it Frisch}, 1999].
An early statistical analysis of the distribution of diffuse and molecular
interstellar clouds concluded that over its lifetime,
a typical disk star such as the Sun
would have encountered over 16 dense interstellar clouds
with radii $\geq$3 pc and density $n_{\rm H}>$10$^3$ cm$^{-3}$,
with more frequent encounters with clouds of lower densities 
[{\it Talbot and Newman}, 1977].  An encounter with a molecular cloud would produce severe
changes in the heliosphere and accretion of ISM onto solar system surfaces.
Dust grains would accrete directly
onto outer solar system surfaces, such as on moons, comets, and Kuiper
Belt objects.  The number of interstellar
dust grains successfully deposited onto these surfaces provides a record of the
combination of the solar galactic environment and activity cycles.
The ratio of large to small dust grains in the deposits provides
a measure of stellar activity cycles.
In addition, cometary surfaces will accrete interstellar dust and the size
distribution of accreted grains will be a function of the orbital
parameters.

The distribution of nearby interstellar material ({\it d}$<$30 pc)
is highly asymmetric; there is an
an order of magnitude more material within 30 pc of 
the Sun in the galactic-center hemisphere, from which the 
material is flowing, than is found in the anticenter hemisphere.
This distribution is also skewed, so that nearby stars at high galactic latitudes
have little foreground ISM ({\it N}(H)$\sim$10$^{18}$ cm$^{-2}$), while stars
at low galactic latitudes ({\it b}$<$--30$^\circ$) 
generally show more nearby gas ({\it N}(H)$\sim$10$^{18.6}$ cm$^{-2}$).

A body of observations indicates that 10--15\% of cool diffuse interstellar gas is 
contained in very small dense structure.  
Multiepoch observations of the H$^\circ$ 21-cm line in absorption 
against high-velocity pulsars show
structure with scale -sizes 5--100 AU and inferred
densities of 10$^3$--10$^5$ cm$^{-3}$ [{\it Frail et al.}, 1994].
Both members of binary star systems have been observed
in search of spatial variations in the optical interstellar Na$^\circ$ D lines,
also indicating diffuse cloud density variations on small
scales.  The ubiquitous presence of small ($<$7000 AU) dense ($>$10$^3$ cm$^{-3}$)
structures in cold interstellar gas is supported by the
these optical data [{\it Meyer and Blades}, 1996; {\it Watson and Meyer}, 1996].
Ultraviolet observations
in $\mu$ Cru find that
Zn$^+$ (which is undepleted and traces both the H$^\circ$ and H$^+$ column densities) 
does not vary between the binary components, so that
the small-scale variations in Na$^\circ$ and other neutrals 
do not represent changes in the total amount of
material present.  
Pockets of cooler and denser material, with enhanced recombination,
would explain the neutral enhancements [{\it Lauroesch et al.}, 1999].
Limits on the density $n_{\rm H}<$50 cm$^{-3}$, for temperature
$\sim$100 K, were derived from the absence of C$^\circ$ fine-structure lines.
The pressure equilibrium of these tiny dense structures is not understood 
because the geometry is unknown, but they would be pressure equilibrium in
two-component systems where the second component is somewhat warmer
than the tiny cold structures [{\it Heiles}, 1997].

Small-scale high-density ionized components
are found near large loops of radio continuum emission (such as Loop I) and are consistent
with postshock, radiatively cooled gas with
electron density $\sim$500 cm$^{-3}$ and sizes $\sim$20 AU
[{\it Heiles}, 1997].

In the LISM cloud complex, observations of the nearest stars indicate
that about one velocity component is seen per 1.7 pc; toward
$\alpha$ Aql at 5 pc, three clouds are seen [{\it Ferlet et al.} 1986]
and toward Sirius at 2.7 pc, two clouds are seen [{\it Bertin et al.}, 1995].
In the direction of $\epsilon$ CMa, which samples the same nearby
gas as Sirius, the second interstellar cloud is slightly cooler
({\it T}=3800 K versus {\it T}=7200 for the LIC) 
and somewhat denser (electron density $n$({\rm e$^-$})=0.46 cm$^{-3}$)
than the LIC [{\it Gry et al.}, 1995].  The second cloud toward $\epsilon$ CMa, 
with log column density log $N$({\rm H$^\circ$})=16.88 cm$^{-2}$,
would be impossible to observe in distant regions unless the
cloud is well separated in velocity from other material in the 
sight line.  The ISM (Fe$^+$, Mg$^+$, D$^\circ$) in the spectrum 
of $\alpha$ Cen, 1.3 pc from the Sun
(and near the LSR upwind direction of the LISM flow),
is not moving with the LIC cloud velocity, suggesting
the LIC has a boundary within 10,000 AU of the Sun
in this direction (see next section).

The bulk of the LISM cloud complex will be flowing past the
solar location during the next 10$^6$ years, with the first
transition of a cloud ``boundary'' within 3000 years if the LIC and
solar velocities are correctly identified.  Therefore
a worthy question is whether small dense structures, similar to
those found in cool ISM, might be embedded in the LISM gas.
This question cannot be answered by existing observations.
Limits on the presence of small dense structures can be placed
by assuming that the LISM is a fractal cloud, within the confines
of its overall morphology 
(e.g., a superbubble shell expanding from the Scorpius-Centaurus
Association).
The maximum density contrast expected for a fractal medium is
$\sim$10$^4$ [{\it Elmegreen}, 1998], which when compared to the $n_{\rm H}$$\sim$0.3
cm$^{-3}$ of the LIC gives a maximum density for a fractal component of 
$n_{\rm H}$$\sim$300 cm$^{-3}$ (where $n_{\rm H}$ is the cloud density).  
Also, $n_{\rm H}$ $\times$ $L$ $<$ 10$^{19}$,
where $L$ is the length of the fractal structure, and the upper limit is set by the approximate
maximum column density of H$^\circ$ through the
LISM gas complex in the upwind direction.  
With this constraint, the upper limit on the sizes of the densest
possible fractal structures
in the LISM would be $L$$<$0.01 pc.  Lower density structures
could be larger.  An encounter with a dense
structure cannot be ruled out observationally, and
a structure of this size,
moving at $\sim$15 km s$^{-1}$, would pass over the Sun in less than 
700 years.  Such a cloud would shrink the heliosphere radius by an order of magnitude.

\section{Velocity Gradient in LISM\label{velocity}}

%Interstellar gas within 30 pc of the Sun clearly exhibits
%a bulk flow past the Sun, with a dispersion of 
%$\sim$2 km s$^{-1}$ about this bulk flow velocity (see below).

The velocity of the interstellar gas immediately surrounding the solar system 
(the LIC) is found from observations of He$^\circ$ within the solar system
(Table 1) [{\it Witte et al.}, 1996; {\it Flynn et al.}, 1998]. 
However, the bulk flow of nearby ISM differs slightly and can be found from the ensemble of 
observations of interstellar gas in nearby stars, where a dispersion of
$\sim$2 km s$^{-1}$ about the bulk flow velocity is found (see below).
Bulk cloud motions derived from optical interstellar absorption lines
are subject to uncertainties because of line weakness in some
directions and higher column densities
in the galactic center hemisphere where individual clouds
usually are not spectrally resolved.
Observations of the nearest star $\alpha$ Cen show several velocity
components (interstellar clouds) within 1.4 pc of the Sun in this direction, and limits on gas
at the LIC velocity place the boundary of the LIC in this
direction at less than 10,000 AU [{\it Landsman et al.}, 1984, 1986; 
{\it Lallement et al.}, 1995; {\it Linsky and Wood}, 1996].

The bulk motion of the LISM cloud complex 
has been determined by observations of interstellar Ca$^+$ absorption lines in nearby stars (see references of F95).  
The average LISM flow, in the velocity rest frame of the Sun, 
corresponds to an inflow velocity --26.8 km s$^{-1}$ from the position
{\it l}=6.2$^\circ$, {\it b}=+11.7$^\circ$ (galactic coordinates), without distinguishing
any velocity between the upwind and downwind directions.
This corresponds to a bulk LISM LSR motion of --15 km s$^{-1}$,
from the direction {\it l}=344$^\circ$, {\it b}=--2$^\circ$, for the new
Hipparcos value for the solar motion (see
Table 1 for other values).
This average motion for nearby interstellar clouds provides a better description
for the LISM than does the assumption that nearby clouds are at rest in 
the LSR, as can be seen in 
\callout{Figure 1}
and
\callout{Figure 2}.
Data on Ca$^+$ absorption lines in 17 nearby stars show 36 absorption line
components (see http://xxx.lanl.gov/astro-ph/9705231, hereinafter
referred to as F97, for data sources).
%These data were acquired at spectral resolutions of 1--3 km s$^{-1}$.
In Figures \ref{fig-lisw} and \ref{fig-lsr},
only one interstellar absorption Ca$^+$ 
component for each of the 17 stars is plotted in two separate reference
frames.  The abscissa gives the component velocity in the LSR rest frame,
while the ordinate gives the component velocity in the heliocentric
local flow velocity frame.  
In Figure \ref{fig-lisw} the plotted component is selected to be the
component which has the smallest velocity in the local flow
velocity frame.  These 17 components have an average velocity in 
the local flow frame of --0.05$\pm$2.24 km s$^{-1}$ (1$\sigma$ 
dispersion).  In Figure \ref{fig-lsr} the plotted component is selected
to be the component which has the smallest velocity in the LSR;
an average velocity of --5.46$\pm$9.82 km s$^{-1}$ is found.
From these plots it is immediately obvious at least one cloud in front
of these stars is moving at the local flow velocity,
and this flow velocity provides a better description of LISM cloud
kinematics than does the assumption that these clouds are at
rest in the LSR.
 
Small regional deviations from an average local
flow vector have been interpreted as indicating the presence of separate
clouds (e.g., the ``LIC'' versus the ``G'' cloud components)
[{\it Lallement and Bertin}, 1992], perhaps indicating a velocity gradient
The gradient is such that nearby interstellar matter in the
galactic-center hemisphere is moving toward the LIC gas
at a relative velocity of about 3 km s$^{-1}$.
The LIC may represent the deceleration of the leading edge of the
LISM cloud complex as it expands into the hot pressurized
plasma interior to the Local Bubble (F95).
Turbulence, associated with the fractal nature
of the ISM, may also explain the observed velocity dispersion 
of $\pm$2.24 km s$^{-1}$ about the flow velocity.
If the LISM gas is warm, the cloud-cloud velocity differences are subsonic.

These velocities and dispersions can be compared with the average
mass-weighted LSR velocity of molecular clouds found in the
solar vicinity, of 2.9 $\pm$0.6 km s$^{-1}$, where the uncertainty
represents 1$\sigma$ (F97).  In contrast, it has been shown that
globally, diffuse clouds with enhanced abundances of the refractory
Ca have much larger velocity dispersions (average velocity 0.9$\pm$11.3 km s$^{-1}$ in the LSR) [{\it Vallerga et al.}, 1993] than do cool clouds, so the general
property of rapidly moving warm clouds with enhanced refractory abundances
is a global characteristic of the ISM.  Given the relatively rapid space motions of diffuse clouds
with enhanced refractory abundances, and the large dimensions associated
with evolved superbubbles, it is not surprising that the 
Sun is currently located in warm material with enhanced refractory abundances.
The LISM bulk motion originates in a region near the
center of the H$^\circ$ 21-cm shell forming the
Loop I superbubble ({\it l,b}=320$^\circ$,+5$^\circ$) [{\it Heiles}, 1998],
supporting the view that this material represents an outflow
from the Scorpius-Centaurus Association.

\section{Interstellar Micrometeorites\label{micrometeorites}}

Interstellar micrometeorites with masses $\sim$10$^{-6.5}$ g have 
been detected by
%AMOR 
radar measurements of ionospheric ion trails; these micrometeorites have inflow directions
preferentially concentrated in the southern ecliptic hemisphere 
[{\it Taylor et al.}, 1996; {\it Baggaley}, this issue;
{\it Landgraf et al.}, this issue].
The total mass of the
detected micrometeorites is a significant fraction of the 
dust mass in the LIC (for particle velocities of 25--50 km s$^{-1}$).
The distribution of nearby interstellar gas is asymmetric,
and the regions of highest nearby column densities are found in
the southern hemisphere (in both ecliptic and galactic coordinates).  
For example, the total column densities toward the stars $\eta$ UMa (ecliptic
coordinates $\lambda$,$\beta$=150$^\circ$,+48$^\circ$) and
$\alpha$ Gru ($\lambda$,$\beta$=317$^\circ$,--33$^\circ$)
are {\it N}(H)$\sim$10$^{18}$ cm$^{-2}$ and {\it N}(H)$\sim$10$^{19}$ cm$^{-2}$, respectively.
Both stars are located 31 pc from the Sun, but in opposite hemispheres.
Therefore both nearby ISM distribution and the micrometeorite 
fluxes show the same broad spatial asymmetries, suggesting
that the micrometeorites may be spatially related to the
interstellar cloud complex which is flowing past the Sun.
This comparison is not a detailed comparison, since
the effects of the solar motion through space have not been considered.

Large particles ($m_{\rm gr}$$>$10$^{-9}$ g) are decoupled from the 
interstellar magnetic field, so these micrometeorites may originate several
hundred parsecs away (F99).  The parent molecular clouds disrupted
by star formation in the Scorpius-Centaurus Association, are suggested
by one model to be the source of the local interstellar cloud complex flowing past the Sun (F95). 
The observed asymmetry in the micrometeorite flux suggests that these molecular clouds may also be the source of the observed micrometeorities.
An alternative origin for the micrometeorites would be in the atmospheres
of evolved stars, since AGBs are a formation site for SiC presolar grains.
The main argument against an AGB origin is that it does not explain
the observed spatial asymmetry.  An origin in the debris from
Type II supernova would also be tenable since $\sim$10 supernova
explosions have occurred during the 15-Myr evolution period of the
Scorpius-Centaurus Association [{\it De Geus}, 1992].

%\begin{figure}
%        \caption{We use the \LaTeX\ {\tt figure} environment
%        to set figure captions.  Figure captions consist of
%        a paragraph containing several sentences or phrases.}
%\end{figure}

\section{Future Outlook}

In principle, the deposition of interstellar dust grains
on geologically inert surfaces in the solar system will
provide a record of changes in the galactic environment of the
Sun.  The heliosphere will contract and expand as
the properties of the interstellar cloud in which it
is immersed change,
with outer planets, comets, and Kuiper Belt objects more likely to be exposed to pristine
interstellar material than inner planets.
When the solar system is embedded in denser interstellar clouds
than at present, the heliosphere will be smaller than it currently
is, exposing the surfaces of outer objects to
the full flux of interstellar dust grains.  These surfaces
may contain a record of these variations in the interstellar dust flux.
Periods when the heliosphere is
large would provide a solar-wind-modulated mass spectrum.
The Earth's Moon will also provide a record of encounters
with interstellar clouds containing large (radii $>$ 0.1 $\mu$m) grains.
If it were possible to obtain and analyze an unmixed core sample of the
lunar surface, the ratio of large to small interstellar grain, as a function
of time, should record a combination of solar activity and the dust
mass spectrum of the cloud in which the Sun was immersed.  
If a similar core sample could be obtained from a surface in the
outer solar system, a comparison with the lunar record would permit disentangling 
the effects of galactic environment versus solar activity.  
Surfaces of Oort cloud comets will accrete interstellar dust.
Thus the changing galactic environment of the Sun on its journey
through space may be recorded by the deposition of interstellar dust grains on solar
system surfaces.  Accessing this record will enrich our understanding of the physical changes which
may have affected the Earth's climate in the past.

\acknowledgments
The author thanks the Department of Astronomy at the Universtiy of 
California, in Berkeley, for its hospitality during this research.  
This research has been supported by NASA grants NAG5-7007 and NAG5-6405 
to the University of Chicago.  

Janet G. Luhmann thanks Adolf N. Witt and Jonathan D. Slavin for their
assistance in evaluating this paper.

   % Type your acknowledgment text (if any) 
   % immediately after the \acknowledgments 
   % command.  Acknowledgments may be only 
   % one paragraph in length.  
   %
   % Be sure to spell "\acknowledgments" 
   % exactly as it appears above; if you spell 
   % it differently then LaTeX will not recognize 
   % the command and it will not work.  There is 
   % no "end acknowledgments" command.
   %
   % If your manuscript contains only one 
   % acknowledgment, use an \acknowledgment 
   % command to produce a singular 
   % "Acknowledgment" subhead.
   %
   % If your manuscript contains Editor's 
   % acknowledgments, set them in a new 
   % paragraph one line below the regular 
   % acknowledgments, as shown.

%% ------------------------------------------------------ %%

%%REFERENCES
\newpage
\clearpage
% ------------------------------------------------------ %%
%%  REFERENCES
%% ------------------------------------------------------ %%

   % In this document we use \markcite commands to refer
   % to citations, so we must enclose references in a 
   % "references" environment.  

        % Include all references between the 
   % \begin{references} and \end{references} 
      % commands.  Each reference must be preceded 
        % by a \reference command.
        % 
       % Extra spacing before each reference is not 
   % necessary; it will not appear in the final 
   % version and is used here only to make the 
        % examples easier to read.  
        %
       % The type of reference being shown appears 
    % in parentheses (after the percent sign) 
      % for each of the following examples; that 
        % information does not need to appear in 
        % your manuscripts or your input files.
        %
    % References should follow the sample format, 
  % with journal titles and issue numbers in 
        % italic type. 
        %
     % Markup commands have been created for some 
        % of the journals referenced most often (such
        % as \jgr or \grl).  Authors may use these 
        % commands as shorthand rather than type out 
        % the whole journal name.  See the aguguide.tex
        % for a full list of short commands.  Note that
        % a comma is automatically included after each
        % journal listing.

\end{article}

\newpage
%%FIGURE CAPTIONS
\clearpage
\begin{figure}
%%\figurenum{17c}
%%\figurewidth{13pc}
%\special{psfile=fig-lisw.ps hscale=120 }
        \caption{The velocities of Ca$^+$ components in 17 nearby stars
are plotted in the local standard of rest (LSR) versus the local interstellar flow velocity frames.  
Only one component per star is plotted.  The plotted components are 
selected to be the components with the smallest velocity in the local
flow velocity frame. 
These components have average velocities of --0.05$\pm$2.24 km s$^{-1}$.
Figure is from http://xxx.lanl.gov/astro-ph/9705231.
The ``LISW''  label on the ordinate is an abbreviation for the
``local interstellar wind,'' which is the flow of interstellar gas
within the solar system (see Table 1).
\label{fig-lisw}
        }
\end{figure}
\begin{figure}
%\special{psfile=fig-lsr.ps hscale=120 }
        \caption{Same as Figure \ref{fig-lisw}, except that the plotted components are
selected to be the components with the smallest velocity in the LSR.
These components have average velocities of --5.46$\pm$9.82 km s$^{-1}$.
Figure is from http://xxx.lanl.gov/astro-ph/9705231.
\label{fig-lsr}}
\end{figure}

%%TABLES
\clearpage
\mbox{
\begin{table}
\begin{center}
\vspace{5pt}
\caption{Properties of the Interstellar Cloud which Surrounds the Solar System\label{licprop}}
\begin{tabular}{lccc} 
\tableline  
\multicolumn{1}{c}{Item}&\multicolumn{1}{c}{Adopted Values}&&\multicolumn{1}{c}{References}\\
\tableline
\\
$n$(He$^{o}$)&0.015 cm$^{-3}$&&1\\
$N$(H$^{o}$)/$N$(He$^{o}$)&14.7&&2, 3, 4\\
$n$(H$^{o}$+H$^{+}$))/($n$(He$^{o}$+He$^{+}$)&10&&10\\
$n$(H$^{o})$&0.22 cm$^{-3}$&&inferred, 15\\
$n$(H$^{+}$)&0.10 cm$^{-3}$&&5, 6, 7, 12, 15\\
Downstream direction in solar rest frame&&&\\
%multicolumn{2}{l}{Downstream direction in solar rest frame}&&\\
$~~~~~$Ecliptic coordinates&$\lambda$=74.7$^{\circ}$$\pm$1.3${^\circ}$, $\beta$=--4.6$^{\circ}$$\pm$0.7$^{\circ}$&&1\\
&{\it V}=24.6$\pm$1.1 km s$^{-1}$&&\\
Upstream directions&&&\\
$~~~~~$Solar rest frame, galactic coordinates&$l$=2.7$^{\circ}$, $b$=+15.6$^{\circ}$ &&\\
&{\it V}=--24.6$\pm$1.1 km s$^{-1}$&&\\
$~~~~~$LSR rest frame, galactic coordinates&$l$=344$^{\circ}$, $b$=--2$^{\circ}$& &14\\
&{\it V}=--14.7 km s$^{-1}$&&\\
Temperature&6,900 K&&11\\
Turbulent velocity &$\sim$1--1.5 km s$^{-1}$ K&&16\\
%(typically about 1.00--1.5 km s$^{-1}$ in the LISM, e. g. Linsky et al. 1995)
Magnetic field&1.5--6 $\mu$G&&9, 13\\
\\
\tableline
\\
\end{tabular}
\end{center}
\end{table}
}
%%\vspace{0.5in}
Table based on {\it  Frisch et al.} [1999].
The flow of interstellar gas through the solar system is
referred to as the local interstellar wind (LISW).
References are
1,  {\it Witte et al.} [1996], A. N. Witte (private communication, 1999), 
and {\it Flynn et al.} [1998];
2,  {\it Dupuis et al.} [1995];
3,  {\it Frisch} [1995];
4,  {\it Vallerga} [1996];
5,  {\it Slavin and Frisch} [1998];
6,  {\it Gry and Dupin} [1996];
7,  {\it Wood and Linsky} [1997];
8,  present paper, section \ref{intro};
9,  {\it Frisch} [1990] (estimated value);
10,  {\it Savage and Sembach} [1996];
11,  {\it Flynn et al.} [1998];
12,   {\it Lallement and Ferlet} [1997];
13,  {\it T. Linde} (private communication, 1997).
14,  present paper, section 4  
(This value is based on the removal of solar motion 
based on by recent Hipparcos data:
13.4 km s$^{-1}$ toward the direction $l$=28$^{\circ}$, $b$=32$^{\circ}$ 
from the observed LISW heliocentric velocity vector (see text).
Earlier estimates of the solar motion yield different results.
Removing ``standard'' solar motion of 
19.5 km s$^{-1}$ toward the direction $l$=56$^{\circ}$, $b$=+23$^{\circ}$,
instead, gives {\it V}=--24.6 km s$^{-1}$, from $l$=315$^{\circ}$, $b$=--3$^{\circ}$ for the observed LISW heliocentric velocity vector.  
If the ``best'' solar motion of 
16.5 km s$^{-1}$ toward $l$=53$^{\circ}$, $b$=+25$^{\circ}$ had been subtracted instead, the
local standard of rest inflow direction of the local interstellar wind would be
{\it V}=--18.2 km s$^{-1}$ from 
$l$=324$^{\circ}$, $b$=--1$^{\circ}$ [e.g., {\it Frisch}, 1995].)
15,  {\it Puyoo and Jaffel} [1998]; and
16,  {\it Linsky et al.} [1995].  The turbulent velocity is defined such
that the line broadening, $b$ (where full width at half maximum=1.6$\times$$b$) is given by the
root-mean-square of the thermal and turbulent velocities.

\clearpage

\clearpage
\newpage
\begin{figure}[t!]
\vspace*{7in}
\includegraphics{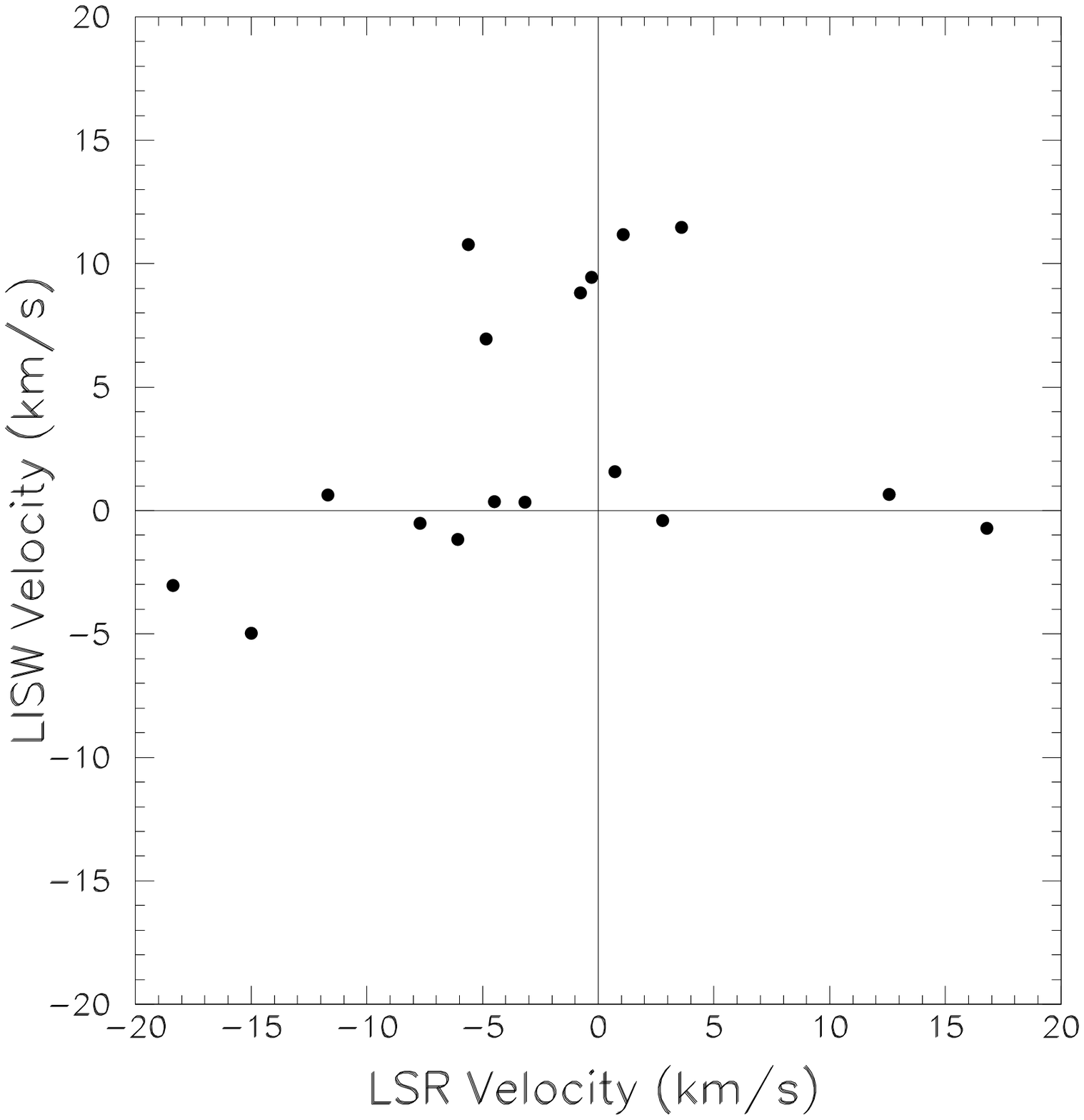}
\end{figure}
\newpage
\begin{figure}[t!]
\vspace*{7in}
\includegraphics{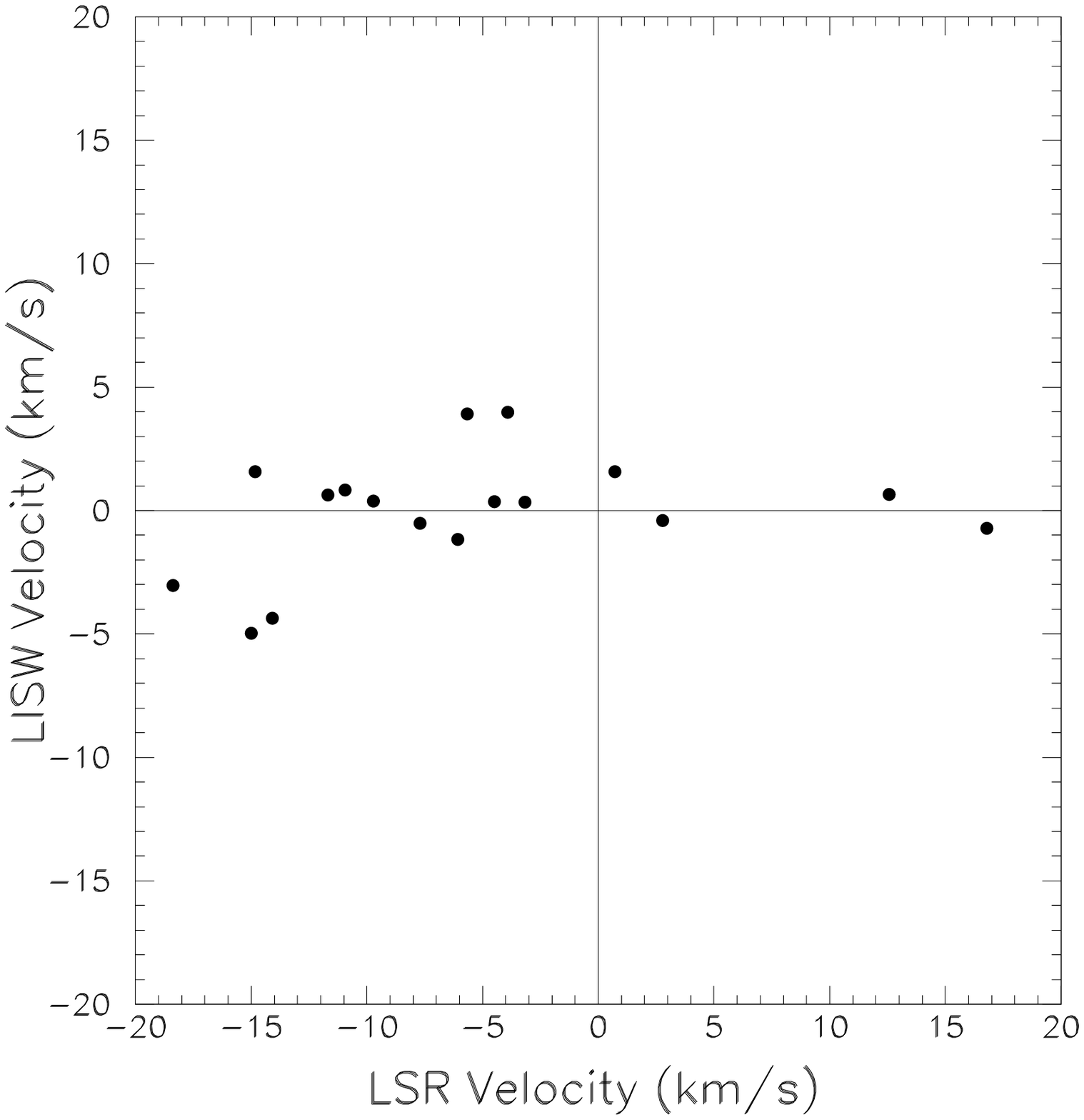}
\end{figure}
\end{document}